\documentclass[twocolumn,showpacs,preprintnumbers,amsmath,amssymb]{revtex4}
\usepackage{amssymb}
\usepackage{graphicx}
\usepackage{dcolumn}
\usepackage{bm}
\usepackage{textcomp}

\begin{document}
\title{Fast creation of conditional quantum gate and entanglement using a common bath only}
\author{Nan Qiu}
\affiliation{State Key Laboratory of Low
Dimensional Quantum Physics, Department of Physics, Tsinghua University, Beijing 100084,
People’s Republic of China}
\author{Xiang-Bin Wang}
 \email{xbwang@mail.tsinghua.edu.cn}
 \affiliation{State Key Laboratory of Low
Dimensional Quantum Physics, Department of Physics, Tsinghua University, Beijing 100084,
People’s Republic of China}
\affiliation{ Jinan Institute of Quantum Technology, Shandong
Academy of Information and Communication Technology, Jinan 250101,
People’s Republic of China}

\begin{abstract}
We propose a scheme to for fast conditional phase shift and creation entanglement of two qubits that interact with a common heat bath. Dynamical decoupling is applied in the scheme so that it works even in the regime of strong interaction between qubits and environment. Our scheme does not
request any direct interaction between the two qubits.
 \end{abstract}
\pacs{03.67.-a, 03.67.Mn, 03.67.Pp} 

\maketitle

\emph{Introduction.} Quantum entanglement and quantum conditional phase shift play the central  role in quantum information processing and quantum commutation \cite{QI},
for example, quantum cryptography with the
Bell theorem \cite{Bell},
quantum dense coding \cite{dc}, quantum teleportation \cite{qt}.
It is therefore an important task to generate entangled states.
Any gate that entangles two qubits, e.g., the conditional phase-shift gate is universal for quantum computation,
when assisted by single-qubit gates \cite{QCEG}. It means that entangling two-qubit gates provide the ability to perform universal quantum computation.
However, it is often very fragile due to
environmental perturbations.
The method of dynamical decoupling (DD) \cite{PDD,CDD,UDD,PUDD,CUDD,QDD,TNDD,TNDD1,MNDD,IMNDD,IMNDD1} can be used to protect the
coherence of qubits in noisy environment, e.g. it can remove the interaction between the system and environment.
DD can also be applied for engineered quantum interaction between qubits and the interaction between qubits and baths \cite{DDEBI1,DDEBI2}.

A smart scheme \cite{EnDB} shows that entanglement between two qubits can be generated if the two qubits interact with a common bath in thermal equilibrium, but not interact directly with each other. This model enhance the usefulness of environment. However, it requests the interaction between qubits and the common bath be weak.
Hence it will cost a long time to prepare entangled state by such a model.

Here we present an efficient method to generate quantum entanglement and make the conditional phase shift gate using only a common heat bath through dynamical decoupling \cite{PDD}. Our method can work in the regime of strong interaction between qubits and environment thus quantum entanglement between qubits can be generated rather fast.
Compared with the existing method \cite{EnDB}, our method can work much more efficiently. In the strong interaction regime,
the coherence of the two qubits would be destroyed rapidly by the environment if there is no designed quantum engineering, e.g., the dynamical decoupling.
Nested UDD \cite{MNDD,IMNDD,IMNDD1} could protect a multi-qubit state, but the nonlocal correlation of qubits is locked by them. Therefore, it cannot generate an entangling two-qubit gate.
Since entangling two-qubit gates result in changing nonlocal correlation of qubits, the control field should reduce the effect on nonlocal correlation of qubits. The control field should reduce the decoherence on the one hand keep the effective two-body Hamiltonian of the two qubits which generates entangle.

The common bath could induce effectively nonlinear couplings in a quantum many-body (multispin) system \cite{LCBGm}. The advantage of the system-bath
coupling is taken by dynamical control so as to realize cooling or heating on a single qubit system \cite{Toc}. Entangled qubits in the common both can
be protected with un-simultaneously DD \cite{DDincom}. Here we simultaneously use UDD on the two qubits \cite{UDD,PUDD} to minimize decoherence,
but make the conditional phase be non-negligible value.
Thus entangling two-qubit gates could be performed fast in strong coupling regime.

\emph{Model.} Two two-level atoms interacting with a common bosonic bath may be described by an extended spin-boson Hamiltonian \cite{CL} $H_{total}=H_S+H_B+H_{int}$ (setting $\hbar=1$ ), where
\begin{eqnarray}
H_S =\frac{\Omega _1}{2}
 \sigma _1^z  + \frac{\Omega _2}{2}\sigma _2^z
\end{eqnarray}
\begin{eqnarray}
H_B=\sum\limits_j {\omega _j a_j^ {\dag}   a_j }
\end{eqnarray}
\begin{eqnarray}
H_{int}=(\sigma _1^z+\sigma _2^z )\left[ {\sum\limits_j {\lambda _j \left( {a_j^ {\dag}    + a_j } \right)}}\right].
\end{eqnarray}
Here $\Omega _i$ is the transition frequency of the $i$th qubit, $\sigma _i^z$ is the Pauli spin operator of the $i$th qubit, and the
environment is represented by a collective bosonic bath with annihilation
(creation) operators $a_j^{(\dag)}$.

\emph{Control pulses.} Consider now $N_d$ instantaneous $\pi$-pulses of $\sigma_1^x$ ($\sigma_2^x$) applied to our system at time $t_{n_1}$ ($t_{n_2}$), with $1\leq n_1\leq N_d$ ($1\leq n_2\leq N_d$). Upon application of one such pulse, one has, in the frame of applied pulses,  $\sigma_1^z\rightarrow -\sigma_1^z$ ($\sigma_2^z\rightarrow -\sigma_2^z$). It hence convenient to introduce the so-called switching function $f_{1(2)}(t)$ , where
\begin{eqnarray}
f_{1(2)}(t)=\sum\limits_{n_{1(2)}}^{N_d}{(-1)^{n_{1(2)}+1}\theta (t-t_{n_{1(2)}})\theta (t-t_{n_{1(2)}+1})},
\end{eqnarray}
with $\theta(t)$ is the Heaviside function.

In the interaction picture this yields
\begin{eqnarray}
 H_I  &=& (\sigma _1^z f_1 (t)+ \sigma _2^z f_2 (t))\times\nonumber\\
 &&\left[ {\sum\limits_j {\lambda _j \left( {a_j^{\dag}  \exp (i\omega _j t) + a_j \exp ( - i\omega _j t)} \right)}  } \right].
 \end{eqnarray}
The closed-form equation for the time-evolution operator (see Appendix A)
takes the simple
form
\begin{eqnarray}\label{eq:seUi}
U\left( t \right) &=& \exp \left[ { - i\int\limits_0^{t} {H_I (t_1 )dt_1}-i\Theta(t)\sigma _1^z \sigma _2^z}\right]\nonumber\\
&=& \exp \left(
 - iH_dt
\right)\exp\left( -iH_{p}t\right),
\end{eqnarray}
where
\begin{eqnarray}
\Theta(t)=-\int\limits_0^t \int\limits_0^{t_1 } && \left( {f_1 \left( {t_1 } \right)f_2 \left( {t_2 } \right) + f_2 \left( {t_1 } \right)f_1 \left( {t_2 } \right)} \right)\times\nonumber\\
&&\sum\limits_j {\left| {\lambda _j } \right|^2 \sin \left[ {\omega _j \left( {t_1  - t_2 } \right)} \right] }{dt_1 dt_2 },
\end{eqnarray}
\begin{eqnarray} \label{eq:Comhp}
H_{p}\equiv\frac{{\Theta(t)\sigma _1^z \sigma _2^z}}{{t}},
\end{eqnarray}
\begin{eqnarray}\label{eq:Comhd}
H_d  \equiv \frac{{\sum\limits_j {(\xi _j (t)a_j^\dag   + \xi _j^ *  (t)a_j )}}}{t},
\end{eqnarray}
\begin{eqnarray}
\xi _j (t) = \int\limits_0^t {ds\lambda _j e^{i\omega _j s} \left( {\sigma _1^z f_1 (s) + \sigma _2^z f_2 (s)} \right)}.
\end{eqnarray}
The evolution of the system, given by Eq.(\ref{eq:seUi}), describes a reservoir-modified i-swap transformation,
and also expresses the decoherence induced by the reservoir. The Hamiltonian $H_p$ generates an entangled gate. However, the Hamiltonian $H_d$ would destroy the coherence.
The control field changes both of them. 
The uncorrelated initial state is given by,
\begin{eqnarray}
\rho_{tot} (0) = \left| \Psi  \right\rangle \left\langle \Psi  \right| \otimes \frac{{e^{ - \frac{{H_B }}{{ k_BT}}} }}{{{\rm {Tr_B}} \left( {e^{ - \frac{{H_B }}{{k_B T}}} } \right)}},
\end{eqnarray}
with $k_B$ denotes Boltzmann’s constant.
We then focuses on the evolution of the reduced state
\begin{eqnarray}\label{eq:ComhdRHO}
\rho _S (t) = {\rm {Tr_B}}(U(t)\rho _{tot} (0)U^\dag  (t)).
\end{eqnarray}
By taking the trace over the field variables of Eq. (\ref{eq:ComhdRHO}) we get
\begin{eqnarray}\label{eq:Comrede}
\rho_S \left( t \right) = \sum\limits_{n,m}^4 {\rho _{n,m} (0)e^{ \left( {i\Theta (t) a_{n,m} - \Upsilon \left( t \right) b_{n,m} } \right)}\left| {\phi _n } \right\rangle \left\langle {\phi _m } \right|},
\end{eqnarray}
where
$\left| {\phi _1 } \right\rangle  \equiv \left| {g_1g_2} \right\rangle ,\left| {\phi _2 } \right\rangle \equiv \left| {g_1e_2} \right\rangle ,\left| {\phi _3 } \right\rangle  \equiv \left| {e_1g_2} \right\rangle ,\left| {\phi _4 } \right\rangle  \equiv \left| {e_1e_2} \right\rangle,
\left( {a_{mn} } \right) = \left( {\begin{array}{*{20}c}
   0 & 2 & 2 & 0  \\
   { - 2} & 0 & 0 & { - 2}  \\
   { - 2} & 0 & 0 & { - 2}  \\
   0 & 2 & 2 & 0  \\
\end{array}} \right)
,
\left( {b_{mn} } \right) = \left( {\begin{array}{*{20}c}
   0 & 2 & 2 & 8  \\
   2 & 0 & 0 & 2  \\
   2 & 0 & 0 & 2  \\
   8 & 2 & 2 & 0  \\
\end{array}} \right)
$, $\left| {g_i} \right\rangle$ is the ground state of the $i$ qubit, $\left| {e_{i}} \right\rangle$ is the excited state of the $i$ qubit.

\begin{figure}
  \includegraphics[width=240pt]{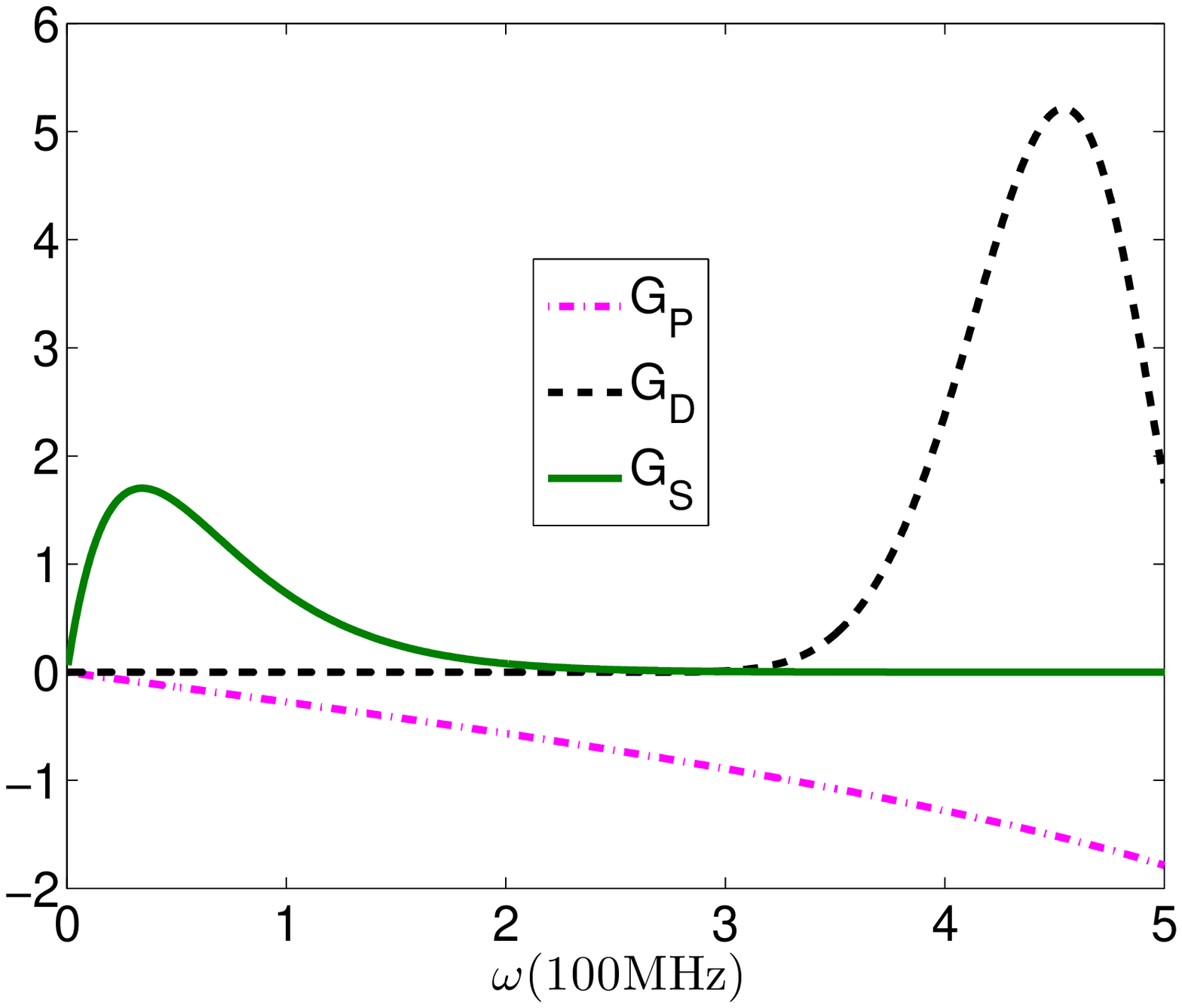}\\
  \includegraphics[width=240pt]{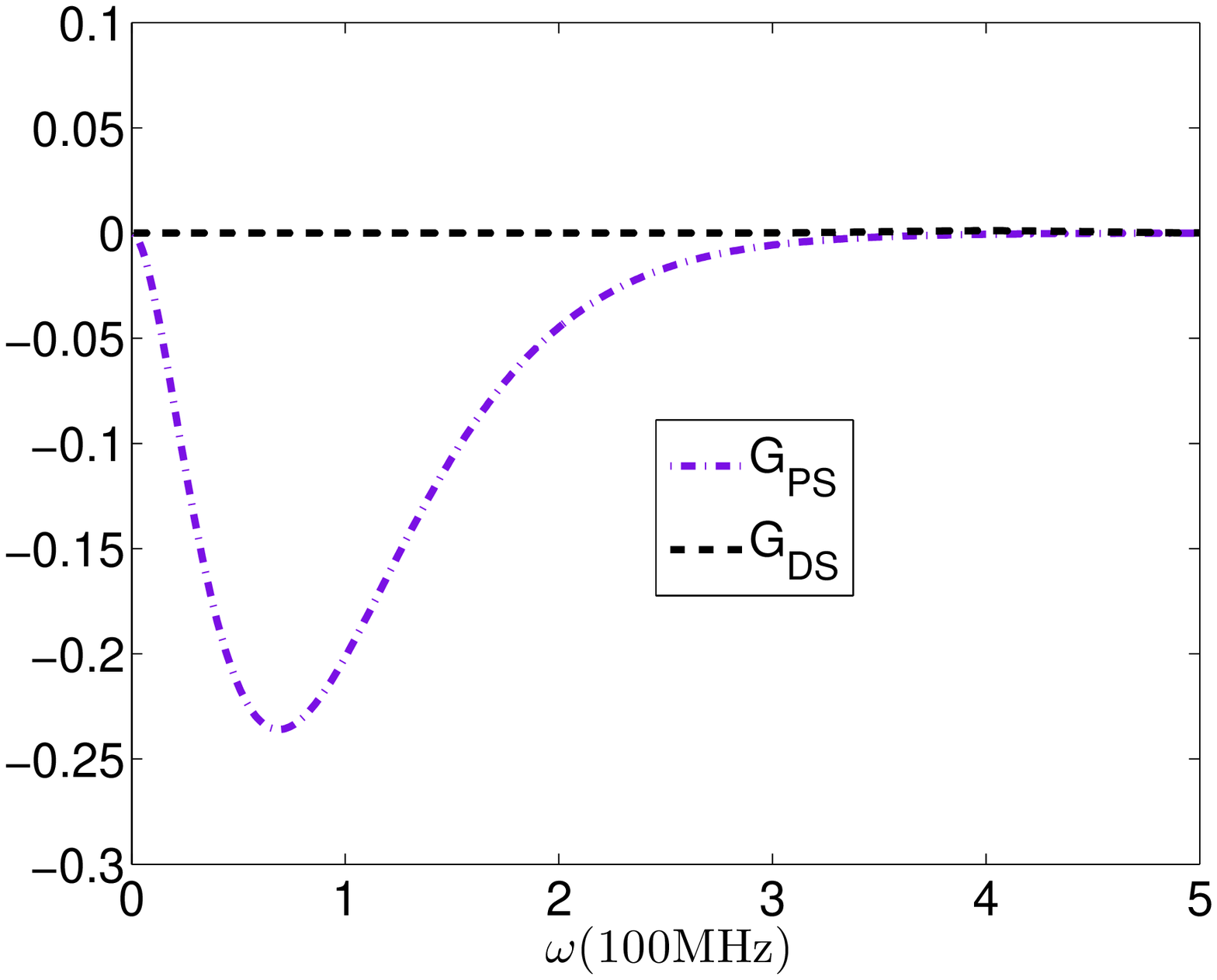}\\
  \caption{(Color online) The functions $G_P \equiv \frac{{10}}
{{\omega ^2 }} [-2M\times\Im _{N_t } (\omega ,\Delta) + \aleph]
$, $G_D \equiv\frac{{10\coth (\frac{\omega }
{{2T}})}}
{{\omega ^2 }}\Re _{N_t } (\omega ,\Delta)$,$G_S \equiv0.1\times J(\omega)$, $G_{PS} \equiv\frac{{J(\omega)}}
{{\omega ^2 }} [-2M\times\Im _{N_t } (\omega ,\Delta) + \aleph]
$, $G_{DS} \equiv\frac{{J(\omega)\coth (\frac{\omega }
{{2T}})}}
{{\omega ^2 }}\Re _{N_t } (\omega ,\Delta)$. Parameters are: $\omega_c=34$MHz; $\eta=135$;
$\Omega_1=10$MHz; $\Omega_2=10$MHz; $T=1$K; $\Delta=16$ns; $M=3$; $N_d=8$.}
\label{DD}
\end{figure}

Created by the common bath, the phase $\Theta(t)$ establishes the nonlocal correlation (entanglement) between the two qubits.

Now we consider to simultaneously use UDD ($f_1(t)=f_2(t)=f(t)$) on the two qubits during the evolution of qubits-bath system form $(l-1)\Delta$ to $l\Delta$, with a number $l$, a time period $\Delta$.
This simultaneous control can on one hand eliminate decoherence on the other hand keep the the nonlocal correlation (entanglement) between the two qubits.
UDD \cite{UDD,PUDD} was originally proposed for suppressing the pure dephasing of a single qubit.
If the pure-dephasing is described by $\sigma_z$-type error (we use standard notation for Pauli matrices), then a UDD sequence
of instantaneous
$\pi$ pulses of the $\sigma_x$ form is applied at
\begin{eqnarray}
t_j  = t\sin ^2 (\frac{{j\pi }}{{2N_d + 2}}), j=1,2,...,N_d,
\end{eqnarray}
with $N_d+1$ pulse intervals during the time period $(0,t]$.  For convenience we also define $t_{N_d+1}=t$.
For odd $N_d$, an additional control pulse is applied at time $t_{N_d+1}$.  Reference\cite{PUDD} proved that such a control sequence can
protect the expectation value of $\sigma_x$ to the $N_d$th order in a universal fashion, irrespective of qubit-environment coupling.
This can be shown by an effective Hamiltonian that only contains even powers of $\sigma_z$.

\emph{Fast generation of quantum entanglement.} After we perform UDD operation above, as the most important quantity, the phase $\Theta$ is given by
\begin{eqnarray}
\Theta (t){\text{ = }}  \int\limits_{\text{0}}^\infty  {d\omega \frac{{J\left( \omega  \right)}}
{{\omega ^2 }}} [-2M\times\Im _{N_d } (\omega ,\Delta) + \aleph].
\end{eqnarray}
Here $J(\omega)$ is the spectrum of standard Ohmic bath
\begin{eqnarray}
 J(\omega)&=& \sum\limits_j {\left| {\lambda _j } \right|^2 \delta \left( {\omega  - \omega _j } \right)}\nonumber\\
&=&\eta \omega e^{-\omega/\omega_c},
\end{eqnarray}
where $\eta$ is the dimensionless
parameter determining the coupling strength between qubit and
bath, $\omega_c$ is the high-energy cutoff value.
The functional $\Im _{N_d }$ is
\begin{eqnarray}
 \Im _{N_d } (\omega ,\Delta) &=&  \omega^2\int\limits_0^\Delta {\int\limits_0^{t_1 } {dt_1 dt_2 f\left( {t_1 } \right)f\left( {t_2 } \right)} } \sin \left[ {\omega _j \left( {t_1  - t_2 } \right)} \right],\nonumber\\
\end{eqnarray}
and
\begin{eqnarray}\label{eq:Comaleph}
&&\aleph  =  - \frac{1}{{{\rm{2}}i}}\int\limits_0^\infty \frac{{J\left( \omega  \right)}}{{ \left( {1 - \cos \omega \Delta } \right)}} \nonumber\\
&&\times\left\{ {\left[ {1 - e^{i\omega \Delta M}  - M\left( {1 - {\mathop{\rm e}\nolimits} ^{i\omega \Delta } } \right)} \right]\left| {f(\omega ,\Delta )} \right|^2  - h.c.} \right\} d\omega,\nonumber\\
\end{eqnarray}
with $M=\frac{t}{\Delta}$.
In the Eq.(\ref{eq:Comaleph}),
\begin{eqnarray}
f(\omega ,\Delta ) = 1 + ( - 1)^{N_d+1 } {\mathop{\rm e}\nolimits} ^{i\omega \Delta }  + 2\sum\limits_{p = 1}^{N_d } {( - 1)^p e^{i\omega \Delta \delta _p } }
\end{eqnarray}
with $\delta_p=\sin^2\frac{\pi\times p}{2\times (N_d+1)}$, is determined by $f(t)$ via the following relation
\begin{eqnarray}
f\left( {\omega ,\Delta } \right) =  - i\omega \int\limits_0^\Delta  {dte^{i\omega t} f\left( t \right)}.
\end{eqnarray}

The entangling gate is performed by the effective Hamiltonian $H_p$.
The concurrence of the two qubits oscillates between zero and one, when the value of $\Theta(t)$ rises.

The decoherence function is given by
\begin{eqnarray}
\Upsilon \left( t \right) = \int_0^\infty  {d\omega \frac{{J\left( \omega  \right)\coth (\frac{\omega }
{{2k_BT}})}}
{{\omega ^2 }}\Re _{N_t } (\omega ,\Delta)},
\end{eqnarray}
with
\begin{eqnarray}
\Re _{N_t } (\omega ,T) = \left| {\frac{{1 - e^{i\omega \Delta M} }}{{1 - e^{i\omega \Delta } }}f\left( {\omega ,\Delta } \right)} \right|^2.
\end{eqnarray}

\begin{figure}
  \includegraphics[width=240pt]{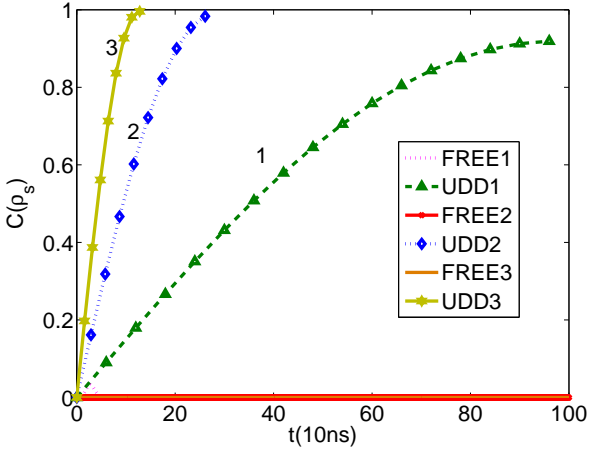}\\
 \caption{(Color online) Comparison of entanglement concurrence for UDD and free evolution. Curve 1: $\eta=1$; $\eta_1=\eta_2=0$; $\omega_c=30$MHz; $T=0.08$mK;
$\Delta$=60ns; $M=16$; $\Omega_1=10$MHz; $\Omega_2=10$MHz; $N_d=9$. Carve 2: $\eta=\eta_1=\eta_2=10$; $\omega_c=\omega_{c_1}=\omega_{c_2}=30$MHz; $T=1$mK;
$\Delta$=29ns; $M=9$; $\Omega_1=10$MHz; $\Omega_2=10$MHz; $N_d=7$.
Carve 3: $\eta=\eta_1=\eta_2=100$; $\omega_c=\omega_{c_1}=\omega_{c_2}=30$MHz; $T=1$K;
$\Delta$=16ns; $M=8$;$\Omega_1=10$MHz; $\Omega_2=10$MHz; $N_d=8$. $\eta_{i}$ is the dimensionless
parameter controlling the coupling between the $i$th qubit and
its individual bath. $\omega_{c_{i}}$ is the high-energy cutoff frequency of the $i$th qubit of individual bath.
}\label{DD}
\end{figure}

The UDD modulation function $f(t)$ changes both the phase (associated with $G_{PS}$) and decoherence (associated with $G_{DS}$). The function $|f(\omega,t)|$ is minimized to its $N_d$-th order in time as shown in Refs. \cite{UDD,PUDD}.  As one can see in the Fig.1, after simultaneous UDD, the peak position of $\Re$ (associated with $G_D$) is moved to a position much larger than the cutoff frequency $\omega_c$ in spectrum density functional $J(\omega)$
 (If the spectrum is not soft \cite{LBUDD1,LBUDD2,LBUDD3}).
 So the overlap  between
  functional $\Re$ and the spectrum density $J(\omega)$ is negligible, which means the decoherence is almost suppressed. For the same reason, $\aleph$ is also almost eliminated.
However, as shown with Ref. \cite{TUDD}, the phase $\Theta$  is quadratic in functional $f(t)$, so that the UDD sequence does not reduce this term. In this case, the functional $\Im$ (associated with $G_P$) takes non-negligible value in low frequency region, while the phase evolution $\Theta(t)$ (associated with $G_{DS}$) is still in action.

The progress for entanglement creation also generates a quantum gate. The entangling gate $U = \exp (i\Theta \sigma _1^z \sigma _2^z )$ refers to conditional phase gate, with $\Theta  \propto M$. If we have the information of the spectrum density which can be detected with DD \cite{MSDD}, we can design the periodic time $\Delta$ and UDD sequence to achieve the entangling gate more effectively.
The general case (common bath and individual baths) is considered in Appendix B.

\emph{Conclusion.} The associated-environment is used to create the entanglement, when simultaneous UDD is applied. The strong coupling between the qubits and the environment is considered. Without UDD, the
decoherence would destroy the correlation between the qubits before the entanglement of two qubits grows up, as shown in Fig.2.
When simultaneous UDD is used, the decoherence is significantly reduced.  The common bath with strong coupling can generate the entanglement in a short time. Within such a short time, the decoherence is negligible. When the common bath is a single-mode harmonic oscillator, our scheme also work. We look forward to developing this scheme in spin squeezing.

\begin{acknowledgments}
We thank Jiangbin Gong for motivating discussion. This work is supported in part by the 10000-Plan of Shandong province and the National High-Tech Program of
China grant No. 2011AA010800 and 2011AA010803, NSFC grant No.
11174177 and 60725416.
\end{acknowledgments}
\section*{APPENDIX}
\subsection{}
To obtain a closed-form expression for the time-ordered unitary operator
\begin{eqnarray}
U(t) = T_ \leftarrow  \exp \left( { - i\int\limits_0^t {H_I (t_1 )dt_1 } } \right),
\end{eqnarray}
we resort to the Magnus expansion of the exponent of $U(t) = \exp(\Omega(t))$. The first few terms of the expansion are
\begin{eqnarray}
 \Omega (t)& =&  - i\int\limits_0^t {H_I (t_1 )dt_1 }  + \frac{1}{2}\int\limits_0^t {dt_1 \int\limits_0^{t_1 } {dt_2 \left[ {H_I (t_1 ),H_I (t_2 )} \right]} }  \nonumber\\
  && +  \cdots.\nonumber\\
\end{eqnarray}
We now take advantage of the remarkable property of bosonic bath operators, namely, that the commutator of the
interaction Hamiltonian at two different times is a C-number function in the bath operators:
\begin{eqnarray}
\left[ {H_I (t_1 ),H_I (t_2 )} \right] =&&  - 2i\left( {f_1 \left( {t_1 } \right)f_2 \left( {t_2 } \right) + f_2 \left( {t_2 } \right)f_1 \left( {t_1 } \right)} \right)\nonumber\\
&&\times\sum\limits_j {\left| {\lambda _j } \right|^2 \sin \left[ {\omega _j \left( {t_1  - t_2 } \right)} \right]\sigma _1^z \sigma _2^z }.\nonumber\\
\end{eqnarray}
Since $\sigma _1^z \sigma _2^z$ commutes with all its powers, the fact that this commutator is a C-number implies that only the first two
terms of the expansion are non-zero.
Now, the closed-form equation for the time-evolution operator takes the simple
form
\begin{eqnarray}\label{eq:Ui}
U\left( t \right) = \exp \left[ { - i\int\limits_0^{t} {H_I (t_1 )dt_1}-i\Theta(t)\sigma _1^z \sigma _2^z}\right],
\end{eqnarray}
where
\begin{eqnarray}
\Theta(t)=\int\limits_0^t {\int\limits_0^{t_1 }  } &&\left( {f_1 \left( {t_1 } \right)f_2 \left( {t_2 } \right) + f_2 \left( {t_1 } \right)f_1 \left( {t_2 } \right)} \right)\nonumber\\
&&\times \sum\limits_j {\left| {\lambda _j } \right|^2 \sin \left[ {\omega _j \left( {t_1  - t_2 } \right)} \right] }{dt_1 dt_2 }.
\end{eqnarray}

\subsection{}

In the case of two qubits are unsymmetrically coupled with their common bath and individual baths, the Hamiltonian in the interaction picture takes the form as

\begin{eqnarray}
H_I  &=& \sigma _1^z f_1 (t)\left[ {\sum\limits_j {\lambda _j \left( {a_j^{\dag}  e^{(i\omega _j t)} + a_j e^{( - i\omega _j t)}} \right)}} \right]\nonumber \\
&&+ \sigma _1^z f_1 (t)\left[{\sum\limits_j {\lambda _{1,j} \left( {a_{1,j}^ {\dag}  e^{(i\omega _{1,j} t)} + a_{1,j} e^{( - i\omega _{1,j} t)}} \right)} } \right]  \nonumber\\
&&+ \sigma _2^z f_2 (t)\left[ {\sum\limits_j {\lambda _j^{'} \left( {a_j^{\dag}  e^ {(i\omega _j t)} + a_j e^ ( - i\omega _j t)} \right)}} \right]\nonumber\\
&& +\sigma _1^z f_2 (t)\left[ {\sum\limits_j {\lambda _{2,j} \left( {a_{2,j}^{\dag}  e^{(i\omega _{2,j} t)} + a_{2,j} e^{ ( - i\omega _{2,j} t)}} \right)} } \right].\nonumber\\
\end{eqnarray}
 The phase is given by
\begin{eqnarray}
  \Theta (t) &=&  - 2\int\limits_0^t {dt_1 \int\limits_0^{t_1 } {dt_2 \sum\limits_j {\lambda _j \lambda '_j {\rm X}(t_1 ,t_2 ,\omega _j )} } }\nonumber\\
   &=&  - 2\int\limits_0^t {dt_1 \int\limits_0^{t_1 } {dt_2 \int\limits_0^\infty  {d\omega } \bar J(\omega ){\rm X}(t_1 ,t_2 ,\omega )} },
\end{eqnarray}
where $
{\rm X}(t_1 ,t_2 ,\omega ) = f(t_1 )f(t_2 )\sin [\omega (t_1  - t_2 )]
$, $\bar J\left( \omega  \right) = \sqrt {J\left( \omega  \right)J^{'} \left( \omega  \right)}$.

The decoherence function is written as
\begin{eqnarray}
 \Upsilon \left( t \right) = \int_0^\infty  {d\omega \frac{{\tilde J\left( \omega  \right)\coth (\frac{\omega }
{{2T}})}}
{{\omega ^2 }}\Re _{N_t } (\omega ,T)},
\end{eqnarray}

where $\Theta _{1,2}= \Theta _{1,3}=2 \Theta$, $ \Theta _{2,4}= \Theta _{3,4}=-2 \Theta$, $ \Theta _{1,4}= \Theta_{2,3}=0$, $\Upsilon_{1,2}=\Upsilon _{1,3}=2\Upsilon$, $\Upsilon_{2,4}=\Upsilon _{3,4}=2\Upsilon$, $\Upsilon_{1,4}=8\Upsilon$, $\Upsilon_{2,3}=2\Upsilon$, $\tilde J_{1,2}\left( \omega  \right)=  J^{'}\left( \omega  \right)+ J_2\left( \omega  \right)$, $\tilde J_{1,3}\left( \omega  \right)= J\left( \omega  \right)+ J_1\left( \omega  \right)$, $\tilde J_{1,4}\left( \omega  \right)= \frac {J\left( \omega  \right)+J^{'}\left( \omega  \right)+2\bar J\left( \omega  \right)+J_1\left( \omega  \right)+J_2\left( \omega  \right)}{4}$, $\tilde J_{2,3}\left( \omega  \right)= J\left( \omega  \right)+J^{'}\left( \omega  \right)-2\bar J\left( \omega  \right)+J_1\left( \omega  \right)+ J_2\left( \omega  \right)$, $\tilde J_{2,4}\left( \omega  \right)= J\left( \omega  \right)+ J_1\left( \omega  \right)$, $\tilde J_{3,4}\left( \omega  \right)= J^{'}\left( \omega  \right)+ J_2\left( \omega  \right)$.



\end{document}